\documentclass[showpacs,prb,preprint]{revtex4}

\usepackage{graphicx}
\usepackage{dcolumn}
\usepackage{amsmath}

\usepackage[T1]{fontenc}
\usepackage{ae}

\newcommand{\leftlead}{\ensuremath{\mathcal{L}}}
\newcommand{\rightlead}{\ensuremath{\mathcal{R}}}
\newcommand{\spacer}{\ensuremath{\mathcal{S}}}

\begin{document}
%
%
\title[Magneto-resistance in ballistic vacuum tunneling]{Bias-voltage
  dependence of the magneto-resistance in ballistic vacuum tunneling:
  Theory and application to planar Co(0001) junctions}

\author{J.~Henk}
\email[Corresponding author. Electronic address:\ ]{henk@mpi-halle.de}
\author{P.~Bruno}

\affiliation{Max-Planck-Institut f\"ur Mikrostrukturphysik\\
  Weinberg 2, D-06120 Halle (Saale), Germany}

\date{\today}

\begin{abstract}
  Motivated by first-principles results for jellium and by
  surface-barrier shapes that are typically used in electron
  spectroscopies, the bias voltage in ballistic vacuum tunneling is
  treated in a heuristic manner. The presented approach leads in
  particular to a parameterization of the tunnel-barrier shape, while
  retaining a first-principles description of the electrodes.  The
  proposed tunnel barriers are applied to Co(0001) planar tunnel
  junctions. Besides discussing main aspects of the present scheme, we
  focus in particular on the absence of the zero-bias anomaly in
  vacuum tunneling.
\end{abstract}



\maketitle

%
%
\section{Introduction}
\label{sec:introduction}
At present, extensive efforts are undertaken to employ the electronic
spin in `magneto-electronic' devices. This aim challenges especially
applied physics, but one is also concerned with model systems of
spin-dependent transport in order to understand the basic
phenomena.~\cite{Maekawa02} Prototypical devices for studies of
ballistic tunneling are planar tunnel junctions (PTJs), which consist
of two magnetic electrodes separated by an insulating spacer. Of
particular interest are the dependencies of the tunnel
magneto-resistance (TMR) on the electronic structure of the leads and
the spacer, on the width of the spacer, and on the bias voltage.

The conductance of a PTJ depends on the density of states (DOS) of the
electrodes and of the tunneling probability of the scattering
channels.~\cite{Slonczewski89} The TMR can then be related to the spin
polarization of the ferromagnetic electrodes.~\cite{Julliere75}
Biasing, which can be viewed as a shift of the chemical potential of
one electrode relative to that of the other, enlarges the range of
energies in which electrons can tunnel through the spacer and
introduces an energy dependence of the electrode spin polarization.

State-of-the-art calculations for spin-dependent tunneling are based
on the very successful density-functional theory
(DFT).~\cite{Gross95a} A bias voltage, however, leads to a
non-equilibrium state, which makes it difficult to apply DFT\@. An
appropriate theoretical description of such a system would require
non-equilibrium Green functions (see, e.\,g.,
Ref.~\onlinecite{Reiss90}).  Therefore, a question arises how one can
maintain the \textit{ab-initio} framework of electronic-structure
calculations, in particular for the leads, but treat the bias voltage
in a feasible manner.

Focusing on spin-dependent ballistic tunneling through PTJs with
finite bias, we investigate in the present work as a simple case
tunneling through a vacuum barrier. The electronic properties of the
electrodes from the spacer can still be computed within spin-polarized
DFT\@. The crucial point is the electrostatic potential in the spacer
region. Guided by first-principles calculations for
jellium~\cite{Lang92} and by theoretical models for surface barriers,
we construct tunnel barriers that show the correct asymptotical
behavior for large spacer thickness. In particular, one of them
compares well with barrier shapes obtained \textit{ab initio} for
jellium. The absence of the zero-bias anomaly (ZBA) in vacuum
tunneling of Co(0001) which was recently found by Ding and
coworkers~\cite{Wulfhekel02a,Ding03a} lends itself support for an
application of the proposed tunnel barrier (for an experimental
investigation of Co PTJs with an oxide barrier, see
Ref.~\onlinecite{LeClair02}). We note in passing that the effect of
interface states on vacuum tunneling in fcc-Co(001) was recently
investigated theoretically.~\cite{Wunnicke02b,Wang03b}

The paper is organized as follows. In Section~\ref{sec:theoretical},
two heuristic ways of constructing a tunnel barrier
(\ref{sec:image-charge-potent} and \ref{sec:superp-surf-barr}) are
motivated. Section~\ref{sec:comp-aspects} deals with computational
aspects of calculations for ballistic tunneling. Results for vacuum
tunneling between Co(0001) electrodes are discussed in
Section~\ref{sec:results-co0001}.

\section{Theoretical}
\label{sec:theoretical}
\subsection{Surface-barrier shapes of metals}
\label{sec:barrier-shapes}
The shape of the surface barrier of a metal was investigated in a vast
amount of publications.  The possibility to calculate accurately
reflected intensities in low-energy-electron diffraction (LEED), which
is a particular surface-sensitive spectroscopy, led to several barrier
models.  Especially at very low energies (VLEED), the shape of the
surface barrier has a considerable effect on the LEED $I(V)$
spectra.~\cite{McRae81} The free parameters that enter its functional
description are fixed by fitting theoretical to experimental data,
e.\,g., to VLEED intensities or to energies of surface and
image-potential states.~\cite{Chulkov99} The latter can be accessed by
inverse or by two-photon photoelectron spectroscopy.~\cite{Grass93}
Note that electronic-structure calculations using the local-density
approximation (LDA) do not reproduce the correct image potential in
the vacuum.

Regarding electron diffraction, the classical electrostatic potential
at a metal surface, with asymptotics $V(z) \approx 1 / (4 z)$, was
first investigated by Mac Coll.~\cite{MacColl39,Units} To avoid the
divergence at the metal surface, Cutler and Gibbons~\cite{Cutler58}
proposed a model potential which interpolated between the (constant)
inner potential $U$ of the metal and the image-charge potential in the
vacuum region. Among the various proposed models, two became the most
popular: the so-called JJJ barrier, named after the inventors Jones,
Jennings, and Jepsen~\cite{Jones84} (see~\ref{sec:superp-surf-barr}
below), and the Rundgren-Malmstr\"om (RM)
barrier~\cite{Rundgren77,Malmstroem80} (for a discussion of JJJ and RM
barriers, see Ref.~\onlinecite{Grass93}).  However, electron scattering is a
dynamical process and these static shapes hold in principle only for
the energy of interest.  An energy-dependent generalization of the JJJ
barrier suggested by Tamura and Feder proved to be successful in
describing the image states at the Pd(110) surface.~\cite{Tamura90a}
Further, the atomic structure at the surface leads to a corrugated
(three-dimensional) surface
potential.~\cite{Tamura83,Lent90,Joly92,Henk93a,Lorenz97}  However,
for most applications the laterally (one-dimensional) and
energetically invariant shapes appear to be sufficient.

\subsection{Construction of the tunnel barrier}
In this Section, we propose two methods of constructing the tunnel
barrier. The first method uses the electrostatic potential of a charge
between two electrodes (\ref{sec:image-charge-potent}), the second
approach is a simple superposition of surface potentials
(\ref{sec:superp-surf-barr}).

\subsubsection{Electrostatic potential between two metal surfaces}
\label{sec:electr-potent-betw}
\begin{figure}
  \centering
  \includegraphics[scale = 1.44]{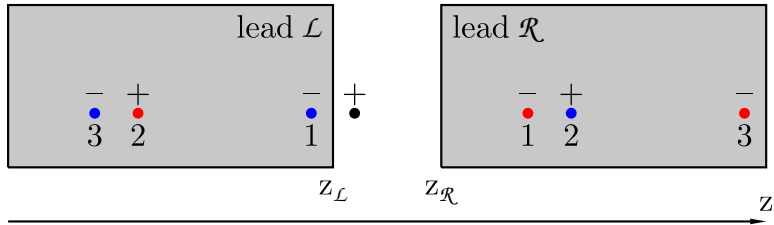}
  \caption{Method of image charges for two metallic leads
    (grey areas). The charge $q = +1$ (black circle) is located
    between the left lead \leftlead\ (with surface at $z_{\leftlead}$)
    and the right lead \rightlead\ (with surface at $z_{\rightlead}$).
    Two series of image charges are obtained by reflection at the
    surfaces, starting with reflection either at $z_{\leftlead}$ (blue
    circles) or at $z_{\rightlead}$ (red circles). Each image charge
    is indicated by the order of reflection ($1, 2, 3, \ldots$) and
    the sign of the charge ($\pm$). Only the first three orders are
    depicted.}
  \label{fig:image}
\end{figure}

Consider a planar tunnel junction with the lead \leftlead\ occupying
the half-space $] -\infty, z_{\leftlead}]$, whereas the lead
\rightlead\ fills $[z_{\rightlead}, \infty[$, with $z_{\leftlead} <
z_{\rightlead}$ (Fig.~\ref{fig:image}).
The electrostatic potential $V_{\mathrm{es}}$ of a charge $q = 1$
between the two semi-infinite metals can easily be obtained by the
method of image charges.~\cite{Units} Because each metal surface acts
as mirror, one has to sum up two infinite series of image-charge
potentials (blue and red circles in Fig.~\ref{fig:image}).  This
procedure results in
\begin{equation}
  V_{\mathrm{es}}(z)
  =
  \frac{1}{4 (z_{\rightlead} - z_{\leftlead})}
  \left[
    2 \gamma
    +
    \Psi\left( \frac{z - z_{\leftlead}}{z_{\rightlead} - z_{\leftlead}} \right)
    +
    \Psi\left( \frac{z_{\rightlead} - z}{z_{\rightlead} - z_{\leftlead}} \right)
  \right],
  \quad
  z \in ] z_{\leftlead}, z_{\rightlead} [.
  \label{eq:1}
\end{equation}
Here, $\gamma \approx 0.577216$ is Euler's constant and $\Psi$ denotes
the Digamma function.~\cite{Abramowitz70} The latter is the
logarithmic derivative of the Gamma function~$\Gamma(z)$, $\Psi(z) =
\mathrm{d} \ln \Gamma(z) / \mathrm{d}z$, with $\Psi(1) = - \gamma$,
$\Psi(z) \propto \ln z$ for $z \to \infty$, and $\Psi(z) \propto
-\gamma - 1/z$ for $z \to 0^{+}$. Obviously, $V_{\mathrm{es}}$
diverges for $z \to z_{\leftlead}$ and $z \to z_{\rightlead}$. It
shows further the well-known asymptotics for the presence of a single
metal.  For example, expanding $V_{\mathrm{es}}$ in a power series
around $z = z_{\rightlead}$ yields
\begin{equation}
  \lim_{z_{\leftlead} \to -\infty} V_{\mathrm{es}}(z)
  =
  -\frac{1}{4 (z_{\rightlead} - z)},
  \quad z \in ]-\infty, z_{\rightlead}[.
\end{equation}
Considering only the first-order approximation for $V_{\mathrm{es}}$,
i.\,e., the direct images of the charge $q$ (labeled $1$ in
Fig.~\ref{fig:image}),
\begin{equation}
  \label{eq:3}
  V_{\mathrm{es}}^{(1)}(z)
  =
  -\frac{1}{4}
  \left[
    \frac{1}{z - z_{\leftlead}}
    +
    \frac{1}{z_{\rightlead} - z}
  \right],
  \quad
  z \in ] z_{\leftlead}, z_{\rightlead} [,
\end{equation}
one sees that $V_{\mathrm{es}}$ represents a higher barrier than
$V_{\mathrm{es}}^{(1)}$ because the images of even order produce an
additional repulsion (`$+$' in Fig.~\ref{fig:image}).

\subsubsection{Image-charge potential as tunnel barrier}
\label{sec:image-charge-potent}
The electrostatic potential between two semi-infinite jellium metals
including a bias voltage was calculated self-consistently within DFT
by Lang.~\cite{Lang92} He found that even with finite bias the
potential in the electrodes is constant a few Bohr radii apart from
the respective surfaces. Further, the divergence of the classical
image-charge potential $V_{\mathrm{es}}$ at $z_{\leftlead}$ and
$z_{\rightlead}$ is bridged over by a smooth interpolating function
which shows the form of a typical LEED-motivated surface barrier
(cf.~\ref{sec:barrier-shapes}).  And last, application of a bias
voltage apparently produces a linear potential drop in the spacer
region (cf.\ Fig.~2b in Ref.~\onlinecite{Lang92}). Guided by these
findings we construct in the following a tunnel barrier by means of
the classical electrostatic potential [eq.~(\ref{eq:1})] and by
LEED-type surface potentials.

\begin{figure}
  \centering
  \includegraphics[scale = 1.44]{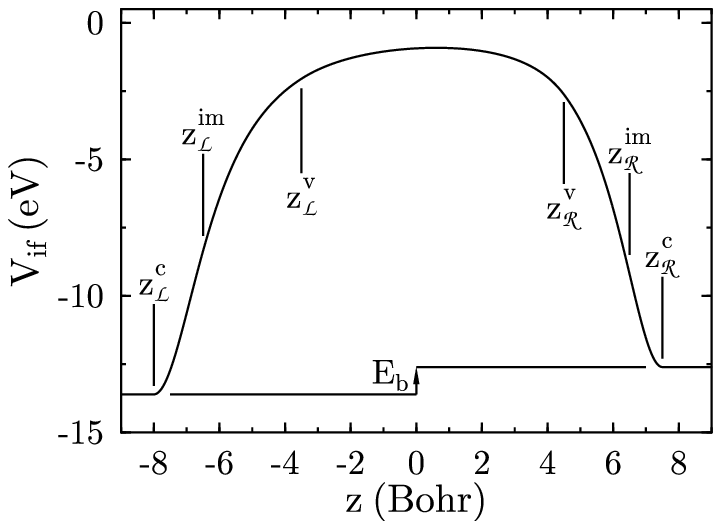}
  \caption{Image-charge potential as tunnel barrier.
    The barrier shape is defined by the parameters as indicated [cf.\ 
    eq.~(\ref{eq:2})]. The Lorentzian shapes extend over the ranges
    $[z_{\leftlead}^{\mathrm{c}}, z_{\leftlead}^{\mathrm{v}}]$ and
    $[z_{\rightlead}^{\mathrm{v}}, z_{\rightlead}^{\mathrm{c}}]$ and
    connect smoothly to the electrostatic potential in
    $[z_{\leftlead}^{\mathrm{v}}, z_{\rightlead}^{\mathrm{v}}]$.  The
    latter comprises the image-charge potential $V_{\mathrm{es}}$
    [eq.~(\ref{eq:1})] as well as the bias potential $V_{\mathrm{b}}$
    [eq.~(\ref{eq:5})].  The inner potentials $U_{\leftlead} =
    13.61~\mathrm{eV}$ and $U_{\rightlead} = 12.61~\mathrm{eV}$ for
    the left and the right lead, resp., determine the bias voltage
    $E_{\mathrm{b}}$ to $+1~\mathrm{eV}$.}
  \label{fig:shape}
\end{figure}

To avoid the divergences of the electrostatic potential
$V_{\mathrm{es}}$, a smooth continuous interpolating function between
the image-charge potential and the inner potentials of the leads,
$U_{\leftlead}$ and $U_{\rightlead}$, is used (The vacuum energy is
taken as energy zero). For this paper we choose a Lorentzian
shape~\cite{Henk93a} but any other reasonable shape can be used, too
(see~\ref{sec:barrier-shapes}).  The interface potential
$V_{\mathrm{if}}$ then reads
\begin{equation}
  \label{eq:2}
  V_{\mathrm{if}}(z)
  =
  \left\{
    \begin{array}{cl}
      -U_{\leftlead}
      &
      z \in ] -\infty, z_{\leftlead}^{\mathrm{c}}]
      \\
      \alpha_{\leftlead}
      \left[1 + \beta_{\leftlead} (z - z_{\leftlead}^{\mathrm{c}})^{2}\right]^{-1}
      + \gamma_{\leftlead}
      &
      z \in [ z_{\leftlead}^{\mathrm{c}}, z_{\leftlead}^{\mathrm{v}}]
      \\
      V(z)
      &
      z \in [ z_{\leftlead}^{\mathrm{v}}, z_{\rightlead}^{\mathrm{v}}]
      \\
      \alpha_{\rightlead}
      \left[ 1 + \beta_{\rightlead} (z - z_{\rightlead}^{\mathrm{c}})^{2} \right]^{-1}
      + \gamma_{\rightlead}
      &
      z \in [ z_{\rightlead}^{\mathrm{v}}, z_{\rightlead}^{\mathrm{c}}]
      \\
      -U_{\rightlead}
      &
      z \in [ z_{\rightlead}^{\mathrm{c}}, \infty [
    \end{array}
  \right..
\end{equation}
The coordinates $z_{\leftlead}^{\mathrm{c}}$, $z_{\leftlead}$ , and
$z_{\leftlead}^{\mathrm{v}}$ specify the positions of the onset of the
interpolating Lorentzian, of the divergence of the potential $V$, and
of the transition to $V$ with respect to \leftlead\ 
(Fig.~\ref{fig:shape}).
They have to be obtained by comparing theoretical results with other
data, e.\,g., surface-state energies, VLEED spectra, etc, for the
surface system (i.\,e., in the limit $z_{\rightlead} \to \infty$).
The parameters $\alpha_{\leftlead}$, $\beta_{\leftlead}$, and
$\gamma_{\leftlead}$ are fixed by the conditions of smooth continuity
in $z_{\leftlead}^{\mathrm{c}}$ and $z_{\leftlead}^{\mathrm{v}}$.
Analogous considerations apply for lead \rightlead.  The potential $V$
in the interior of the spacer can be chosen to incorporate the
electrostatic potential between two metal electrodes,
$V_{\mathrm{es}}$, and the bias voltage as well, as being discussed in
the following.

Bringing two metals so close that electrons can tunnel from metal to
the other aligns the Fermi levels of the two leads. This energy shift
is given by the contact potential $\Phi_{\leftlead} -
\Phi_{\rightlead}$, i.\,e., the difference of the work functions
$\Phi_{\leftlead}$ of \leftlead\ and $\Phi_{\rightlead}$ of
\rightlead. Note that the alignment of the Fermi levels is accompanied
by a shift of the inner potentials, that is, e.\,g., $U_{\rightlead}$
of the semi-infinite system is replaced by $U_{\rightlead} -
\Phi_{\leftlead} + \Phi_{\rightlead}$.  Since $V(z)$ was not specified
explicitly in eq.~(\ref{eq:2}), it can account for the contact
potential and the bias voltage,
\begin{equation}
  \label{eq:4}
  V(z)
  =
  V_{\mathrm{es}}(z)
  +
  V_{\mathrm{b}}(z).
\end{equation}
Here, $V_{\mathrm{es}}$ is the electrostatic potential from
eq.~(\ref{eq:1}) and $V_{\mathrm{b}}$ is the bias, for which a linear
drop over the interface region is assumed:
\begin{equation}
  \label{eq:5}
  V_{\mathrm{b}}(z)
  =
  \left\{
    \begin{array}{cl}
      0
      &
      z \in ] -\infty, z_{\leftlead}^{\mathrm{c}}]
      \\
      E_{\mathrm{b}}
      \frac{z - z_{\leftlead}^{\mathrm{c}}}{z_{\rightlead}^{\mathrm{c}} - z_{\leftlead}^{\mathrm{c}}}
      &
      z \in [z_{\leftlead}^{\mathrm{c}}, z_{\rightlead}^{\mathrm{c}}]
      \\
      E_{\mathrm{b}}
      &
      z \in [z_{\rightlead}^{\mathrm{c}}, \infty [
    \end{array}
  \right..
\end{equation}
This \textit{ansatz} is motivated by the fact that the electric field
is well screened within the electrodes but unscreened within the
vacuum spacer.

\begin{figure}
  \centering
  \includegraphics[scale = 1.44]{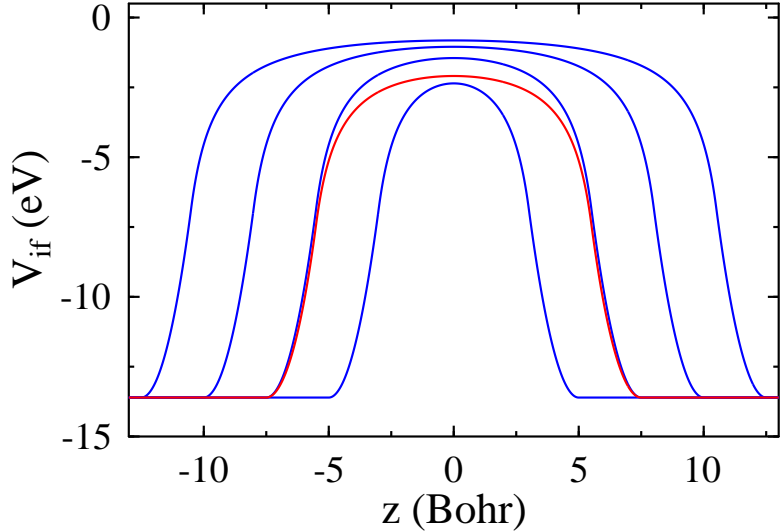}
  \caption{Dependence of the interface potential $V_{\mathrm{if}}$
    [eq.~(\ref{eq:2})] on the lead separation.  The barriers with
    electrostatic potential $V_{\mathrm{es}}$ [eq.~(\ref{eq:1})] are
    shown for separations of $10$, $15$, $20$, and $25$~Bohr radii
    (blue) at zero bias ($E_{\mathrm{b}} = 0$~eV).  In addition, a
    barrier with first-order approximation potential
    $V_{\mathrm{es}}^{(1)}$ [eq.~(\ref{eq:3})] is shown for $15$~Bohr
    radii separation (red). The inner potentials of the leads are
    equal ($U_{\leftlead} = U_{\rightlead} = 13.61$~eV).}
  \label{fig:series}
\end{figure}

\begin{figure}
  \centering
  \includegraphics[scale = 1.44]{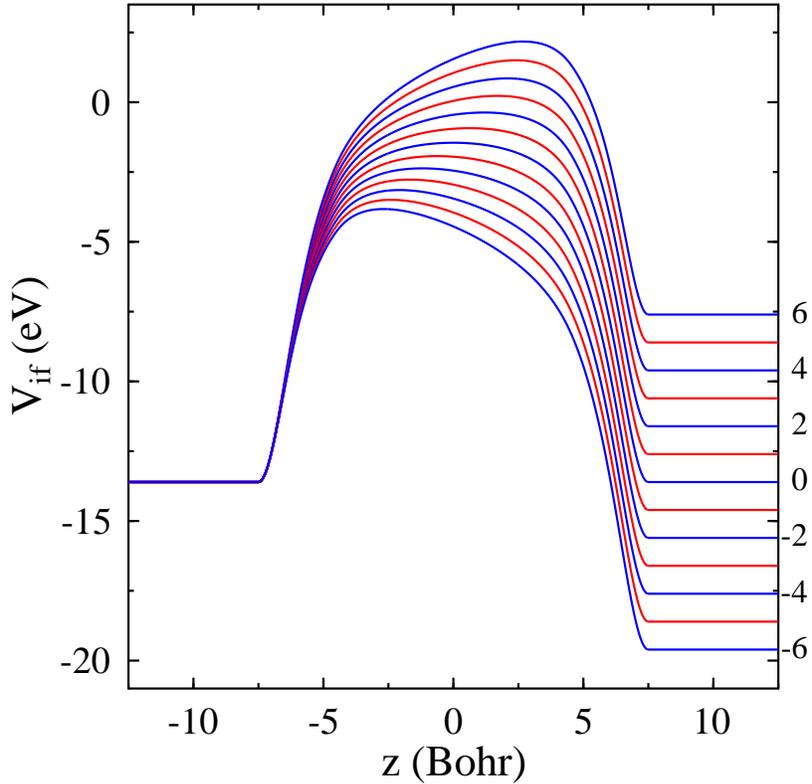}
  \caption{Dependence of the interface potential $V_{\mathrm{if}}$
    [eq.~(\ref{eq:2})] on the bias.  For a lead separation of $15$
    Bohr radii, the right lead \rightlead\ is biased from
    $-6~\mathrm{eV}$ to $+6~\mathrm{eV}$ (as indicated on the right;
    alternating blue and red lines; $U_{\leftlead} = U_{\rightlead} =
    13.61$~eV).}
  \label{fig:series2}
\end{figure}

Figure~\ref{fig:series} presents a series of tunnel barriers in
dependence of the lead separation $z_{\rightlead}^{\mathrm{c}} -
z_{\leftlead}^{\mathrm{c}}$.
The heights of the barriers increase with separation (for an
experimental estimation of the barrier height \textit{vs} distance,
see Ref.~\onlinecite{Wulfhekel02a}). As already mentioned, taking the
first-order approximation $V_{\mathrm{es}}^{(1)}$ instead of
$V_{\mathrm{es}}$ leads to a reduced height (cf.\ the red line for
15~Bohr radii lead separation).  The shape dependence on the bias is
addressed in Fig.~\ref{fig:series2}.
For rather large bias, the linear potential drop in the interface
region can be clearly retrieved. We note in passing that the present
construction produces potential shapes that compare qualitatively well
with those obtained \textit{ab initio} for jellium by
Lang.~\cite{Lang92}  Further, a similar approach was recently used to
explain the Stark shifts of surface states in scanning tunneling
spectroscopy.~\cite{Limot03}

One advantage of the present approach is that the height of the tunnel
barrier is automatically adjusted in dependence on the lead separation
and on the bias voltage.  Further, the barrier shape shows the correct
image potential asymptotics for large lead separation [cf.\
eq.~(\ref{eq:3})]. In turn, the approach should not be applied for too
small separations because the barrier shape would significantly differ
from the interface potential which would be obtained from a
self-consistent calculation for a narrow tunnel junction. This,
however, could possibly be compensated by adjusting the parameters
$z_{\leftlead}^{\mathrm{c}}$, \ldots, $z_{\rightlead}^{\mathrm{c}}$
not for the semi-infinite system but for the narrow junction.

\subsubsection{Superposition of surface barriers}
\label{sec:superp-surf-barr}
For large lead separations and small bias voltages, the probability of
electrons to tunnel from one lead to the other is very small. Hence,
in a self-consistent calculation for a tunnel junction, the tunnel
barrier appears to be almost exclusively determined by the electron
density of the respective lead and not significantly influenced by
that of the other lead. This consideration might lead one to construct
a tunnel barrier by superposition of the respective surface barriers,
\begin{equation}
  \label{eq:6}
  V_{\mathrm{if}}(z)
  =
  V_{\leftlead}(z)
  +
  V_{\rightlead}(z).
\end{equation}
where $V_{\leftlead}$ and $V_{\rightlead}$ are the surface potentials
of the respective leads. Taking JJJ barriers,~\cite{Jones84} one
arrives at
\begin{equation}
  \label{eq:jjjL}
  V_{\leftlead}(z)
  =
  \left\{
    \begin{array}{cl}
      \frac{1}{4 (z_{\leftlead} - z)}
      (1 - \exp[\lambda_{\leftlead} (z_{\leftlead} - z)])
      &
      z \in [ z_{\leftlead}, \infty[
      \\
      -\frac{U_{\leftlead}}{\alpha_{\leftlead} \exp[\beta_{\leftlead} (z_{\leftlead} - z)] + 1}
      &
      z \in ] -\infty, z_{\leftlead}]
    \end{array}
  \right.
\end{equation}
for the surface barrier of \leftlead.  The values of
$\alpha_{\leftlead}$ and $\beta_{\leftlead}$ are determined by
requiring smooth continuity at $z = z_{\leftlead}$.  Because
$U_{\leftlead}$ is known from the self-consistent calculation for the
surface system, $z_{\leftlead}$ and $\lambda_{\leftlead}$ remain as
the only parameters to be adjusted. For the surface potential of
\rightlead\ one obtains an analogous form.

The simple superposition of surface barriers appears to be problematic
for hetero-junctions or biased junctions. In both cases, the relative
energy shift of one electrode, say \rightlead, results in a finite
potential which extends into the entire other electrode \leftlead.
This is due to the fact that the surface potential of \rightlead\ 
extends infinitely far into \leftlead.  One way to overcome this
problem is to take the bias only as an energy shift in the interior of
\rightlead, that is, to replace the inner potential $U_{\rightlead}$
by $U_{\rightlead} - E_{\mathrm{b}}$.

\begin{figure}
  \centering
  \includegraphics[scale = 1.44]{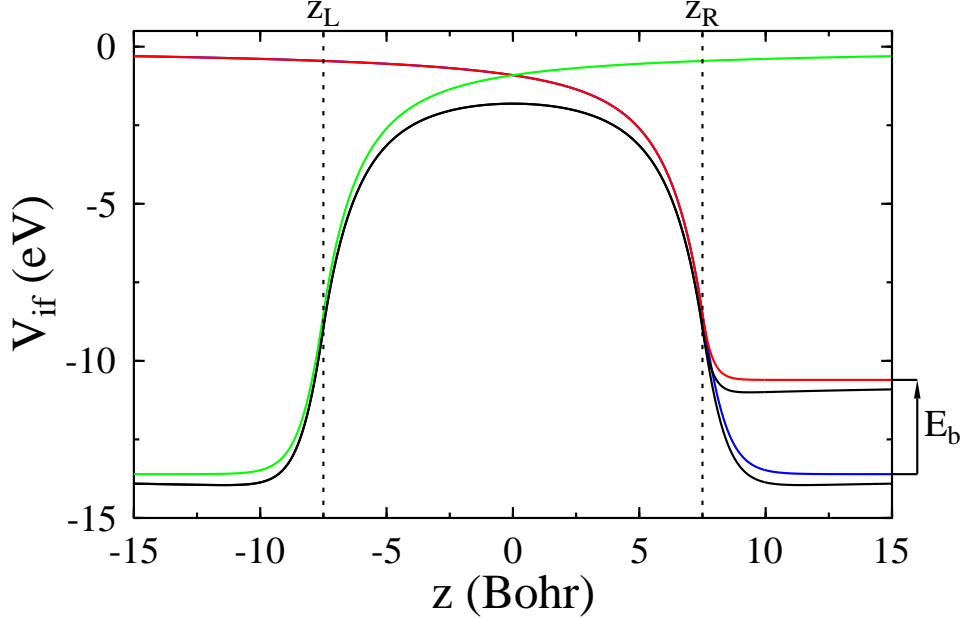}
  \caption{Formation of an interface barrier by superposition of
    surface barriers. Without bias, two surface barriers of JJJ type
    (\leftlead\ green and \rightlead\ blue) are superposed to yield
    the interface barrier (black). The inner potentials are equal
    ($U_{\leftlead} = U_{\rightlead} = 13.61~\mathrm{eV}$).  The
    application of a bias shifts the inner potential of \rightlead\ 
    (red; $E_{\mathrm{b}} = 3~\mathrm{eV}$, cf.\ the arrow) and
    results in the other tunnel barrier (also black).  The positions
    of the image potential divergences are $z_{\leftlead} = -7.5\,
    a_{0}$ and $z_{\rightlead} = 7.5\, a_{0}$ (Bohr radii),
    respectively (cf.\ the vertical dotted lines),
    $\lambda_{\leftlead} = \lambda_{\rightlead} = 1.25 / a_{0}$
    [eq.~(\ref{eq:jjjL})]. The vacuum level is taken as energy zero.}
  \label{fig:barriershape}
\end{figure}

Figure~\ref{fig:barriershape} shows a superposition of JJJ surface
barriers.
The inner potential of \rightlead\ is shifted (cf.\ the arrow) by
$E_{\mathrm{b}} = 3~\mathrm{eV}$. Apparently, the barrier shape does
not change significantly with bias in $[z_{\leftlead},
z_{\rightlead}]$, in contrast to the former construction
(Fig.~\ref{fig:series2}). In particular, the linear potential drop is
not observed.

\subsubsection{R\'{e}sum\'{e}}
\label{sec:resume}
The two construction recipes result in tunneling barriers with
different features. While the more elaborate one [eq.~(\ref{eq:2})]
produces barrier shapes which are qualitatively close to those
obtained from first-principles for jellium,~\cite{Lang92} the barriers
of the superposition approach [eq.(\ref{eq:6})] lack most of these
important features. In particular, the linear potential drop in the
spacer region is missing.  Both approaches can easily be extended to
energy-dependent and corrugated (three-dimensional) tunnel barriers.

Wang and coworkers obtained the shape of the tunnel barrier by
matching two surface systems that were calculated for equal but
opposite shifts of the work functions.~\cite{Wang03b} The resulting
bias-dependent barriers agree well with that shown in
Fig.~\ref{fig:series2}.

\subsection{Computational aspects of ballistic tunneling}
\label{sec:comp-aspects}
For the ballistic-tunneling calculations we applied the layer-KKR
(Korringa-Kohn-Rostoker) approach of MacLaren and
coworkers~\cite{MacLaren99} which is based on the Landauer-B\"uttiker
result for the tunnel conductance.~\cite{Buettiker86} At a given
energy $E$ and in-plane crystal momentum $\vec{k}_{\parallel}$, one
computes the Bloch states $n_{\leftlead}$ and $m_{\rightlead}$ of the
electrodes \leftlead\ and \rightlead\ and classifies them with respect
to their propagation direction: to the right ($+$) or to the left
($-$). The scattering matrix $S$ of the spacer \spacer\ is first
computed in a plane-wave basis using LEED algorithms (like
layer-doubling and layer-stacking; see for example
Ref.~\onlinecite{Henk01c}) and subsequently expressed in terms of the
scattering channels, i.\,e., in the Bloch-state basis.  The
transmission $T(E_{\mathrm{t}}, \vec{k}_{\parallel})$ at the tunnel
energy $E_{\mathrm{t}}$ is then a sum over all pairs of Bloch states
that are incident in $\leftlead$ and outgoing in $\rightlead$,
\begin{equation}
  \label{eq:transmission}
  T(E_{\mathrm{t}}, \vec{k}_{\parallel})
  =
  \sum_{n_{\leftlead}, m_{\rightlead}}
  \left|
  S_{n_{\leftlead} m_{\rightlead}}^{++}(E_{\mathrm{t}}, \vec{k}_{\parallel})
  \right|^{2}.
\end{equation}
The tunnel conductance $G(E_{\mathrm{t}})$ is obtained by summing over
the two-dimensional Brillouin zone (2BZ),
\begin{equation}
  \label{eq:conductance}
  G(E_{\mathrm{t}})
  =
  G_{0}
  \sum_{\vec{k}_{\parallel} \in \mathrm{2BZ}}
  T(E_{\mathrm{t}}, \vec{k}_{\parallel}).
\end{equation}
Here, $G_{0} = e^{2} / h$ is the quantum of conductance which equals
$2 \pi$ in atomic units.~\cite{Units} Adaptive mesh refinement
provides an efficient method to obtain accurate and well-converged 2BZ
sums, in particular, if small parts of the 2BZ contribute
significantly to the conductance.~\cite{Henk01b}

With a bias voltage applied, electrons can tunnel from occupied states
of one lead into unoccupied states of the other lead. The total
conductance is then obtained by integrating over the energy interval
given by the Fermi energies $E_{\mathrm{F}}$ of the electrodes. The
averaged conductance thus reads
\begin{equation}
  \label{eq:7}
  G_{\mathrm{av}}
  =
  \frac{1}{\left| E_{\mathrm{F}\leftlead} - E_{\mathrm{F}\rightlead} \right|}
  \int_{\min(E_{\mathrm{F}\leftlead}, E_{\mathrm{F}\rightlead})}^{\max(E_{\mathrm{F}\leftlead}, E_{\mathrm{F}\rightlead})}
  G(E_{\mathrm{t}})
  \,\mathrm{d}E_{\mathrm{t}}.
\end{equation}
The tunnel magneto-resistance (TMR) $\rho$ is defined as the asymmetry
of the (averaged) conductances for parallel (P) and antiparallel (AP)
alignment of the electrode magnetizations,
\begin{equation}
  \label{eq:8}
  \rho
  =
  \frac{G_{\mathrm{av}}(\mathrm{P}) - G_{\mathrm{av}}(\mathrm{AP})}{G_{\mathrm{av}}(\mathrm{P}) + G_{\mathrm{av}}(\mathrm{AP})}.
\end{equation}

To treat in practice the bias voltage we proceed as follows. First,
self-consistent electronic-structure calculations for the
semi-infinite leads \leftlead\ and \rightlead\ generated muffin-tin
(MT) potentials of the bulk, of the surface, and in the vacuum region.
The MT zeroes were taken as inner potentials $U_{\leftlead}$ and
$U_{\rightlead}$, respectively (the MT zero is the constant potential
in the interstitial region).  For each of the leads, the MT potentials
in the vacuum region were replaced by a smooth surface barrier. The
parameters of the latter were fixed by requiring that the spectral
densities in the surface layers were as close as possible to that of
the original self-consistent calculation.  Here, the focus laid in
particular on the energy range used in the subsequent tunneling
calculations and on surface states. For Co(0001), an important feature
is the energy of the majority surface state at $\vec{k}_{\parallel} =
0$ (cf.\ Fig.~2b in Ref.~\onlinecite{Ding03a}; see also
Refs.~\onlinecite{Math01,Braun02,Okuno02}).

Having fixed the barrier parameters, the tunnel junction was built
from the bulk and surface potentials of the two electrodes and the
interface barrier [eqs.~(\ref{eq:2}), (\ref{eq:4}), and (\ref{eq:5})].
The spacer \spacer\ comprises all layers with potentials that differ
from the respective bulk potentials. In example, for a Co(0001) tunnel
junction the first four layers on either side of the smooth tunnel
barrier were used. The bias was taken into account by shifting the
inner potential of one of the leads (muffin-tin zero) and determining
the barrier shape [eq.~(\ref{eq:1})]. The smooth barrier
$V_{\mathrm{if}}$ was treated as a single layer in the
multiple-scattering calculations. Its scattering matrix $S$ was
obtained within the propagator formalism.~\cite{Jepsen71}

As usual for the KKR method, a small imaginary part $\eta$ has to be
added to the energy $E$,~\cite{Weinberger90} leading in general to
complex wavenumbers $k_{\perp}$.~\cite{Slater37,Heine63} Therefore,
the electrode eigenfunctions are no longer true Bloch states but
become evanescent states [$\Im(k_{\perp}) \not= 0$]. Eigenstates
stemming from Bloch states [$\Im(k_{\perp}) = 0$ for $\eta = 0$] show
typically the smallest $\Im(k_{\perp})$ and can therefore be separated
from evanescent states [$\Im(k_{\perp}) \not= 0$ for $\eta =
0$].~\cite{MacLaren99} In the spacer \spacer, the nonzero $\eta$ leads
to damping in addition to the intrinsic one, artificially enhancing
the decay of the conductance with spacer thickness.  Further, the
scattering matrix $S$ is no longer unitary and, hence, the total
current is not conserved.  Therefore, one has to choose $\eta$
carefully in order to produce reliable results. We found that a value
of $\eta = 10^{-4}~\mathrm{eV}$ produces no considerable artefacts.

The Landauer-B\"uttiker approach used here avoids the computation of
the Green function of the complete system, which is in particular
problematic for a non-equilibrium system. Considering the asymptotic
transmission channels (Bloch states), states that are localized at the
barrier do not contribute to the transmission.

Recently, Davis and MacLaren reported on model calculations for
spin-dependent tunneling at finite bias.~\cite{Davis00} In their work,
however, the electronic structure of the Fe electrodes was
approximated by plane waves, whereas the barrier was assumed as
step-like with a linear drop. Although conceptual similar, our
approach goes beyond that work. First, the electrodes are treated on
first-principles level. Second, the barrier shows the correct
asymptotics (for the free surfaces) and, once the shape parameters
being fixed, depends automatically on both lead separation and bias.

\section{Results for C\lowercase{o}(0001)}
\label{sec:results-co0001}
Recently, Ding and coworkers investigated the bias-voltage dependence
of the TMR with a spin-polarized scanning tunneling microscope
(STM).~\cite{Ding03a} In contrast to tunneling through oxide barriers,
they observed no zero-bias anomaly (ZBA), i.\,e., a (rather) sharp
maximum of the TMR at zero bias (see, for example,
Ref.~\onlinecite{Yuasa02}).  With a vacuum barrier replacing an oxide
barrier, the TMR appeared to be almost constant. This finding suggests
that the ZBA is mainly due to imperfections in oxide barriers, rather
than to scattering at magnons and spin excitations (in the leads).
Further, the so-called DOS effect, i.\,e., the energy dependence of
the spin-resolved density of states of the leads, proved to be small
in the case of Co(0001).

The experimental findings of Ding \textit{et al.} were corroborated by
ballistic tunneling calculations for planar Co(0001) junctions as
sketched in Section~\ref{sec:comp-aspects}.  The tunnel barrier was
taken as a superposition of surface barriers
(\ref{sec:superp-surf-barr}). In the present work, we focus on the
more elaborate image-charge potential (\ref{sec:image-charge-potent}).

\begin{figure}
  \centering
  \includegraphics[scale = 0.5]{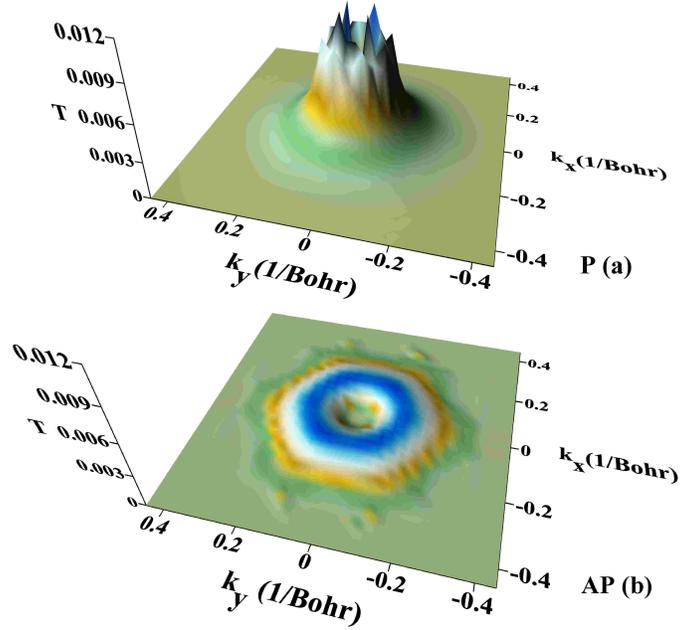}
  \caption{Transmission $T(E_{\mathrm{F}}, \vec{k}_{\parallel})$ of vacuum
    tunneling in Co(0001) for P (a, top) and AP (b, bottom) alignment
    of the lead magnetizations. For $0~\mathrm{eV}$ bias, the lead
    separation was chosen as $7.52$~\AA. The maximum transmission is
    about $0.01$ (P alignment. Both panels share the same scale).
    Note that only the central part of the two-dimensional Brillouin
    zone is displayed.}
  \label{fig:transmission}
\end{figure}

The transmission $T(E_{\mathrm{F}}, \vec{k}_{\parallel})$,
eq.~(\ref{eq:transmission}), depends on the relative orientation of
the lead magnetizations (P and AP), as is shown for $0~\mathrm{eV}$
bias (tunneling at $E_{\mathrm{F}}$) in Fig.~\ref{fig:transmission}.
For the chosen lead separation of $7.52$~\AA, only those Bloch states
with a $\vec{k}_{\parallel}$ in the central part of the 2BZ contribute
significantly to the transmission. The normal component of the
wavevector within the tunnel barrier, $k_{\perp}(z) = \sqrt{2
  [E_{\mathrm{F}} - V_{\mathrm{if}}(z)] - \vec{k}_{\parallel}^{2}}$,
is imaginary and gives rise to strongly evanescent states in the
tunnel barrier for Bloch states with large $\vec{k}_{\parallel}$, and,
thus, to a small transmission. For Bloch states with
$\vec{k}_{\parallel}$ near $\overline{\Gamma}$ (i.\,e.,
$\vec{k}_{\parallel} = 0$), the decay within the barrier is less and
the transmission can be larger. In total, this results in a `focusing'
of $T(E_{\mathrm{t}}, \vec{k}_{\parallel})$ at the 2BZ center.

Both the P and the AP case show minor transmission close to
$\overline{\Gamma}$.  These minima are surrounded by ring-like
structures of increased transmission. The maximum P transmission is
larger than for AP alignment (by a factor of about $10$). But the AP
transmission displays a broader ring compared to the P transmission.

\begin{figure}
  \centering
  \includegraphics[scale = 1.0]{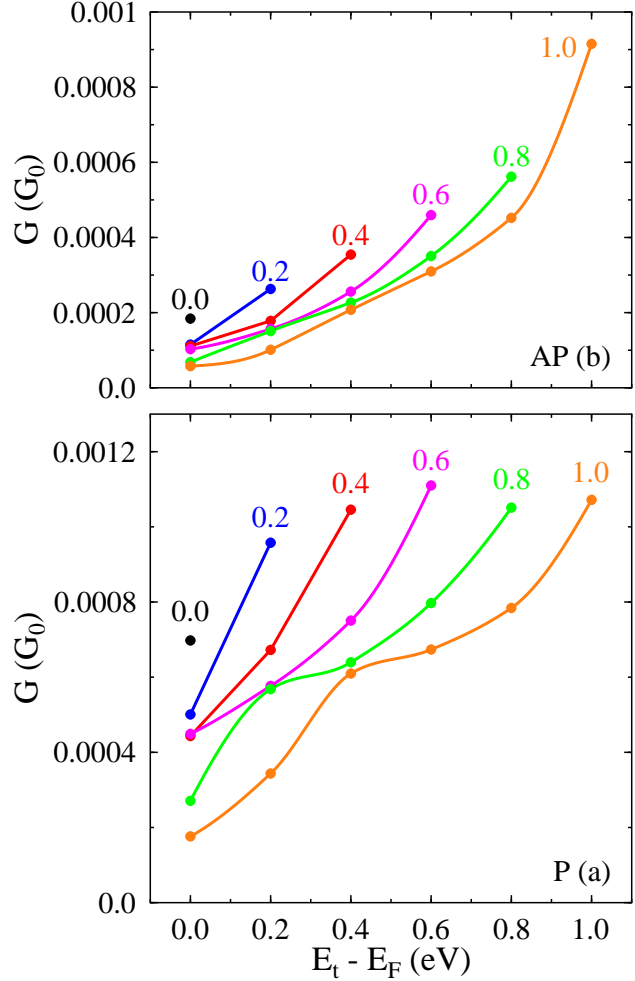}
  \caption{Ballistic conductance $G$ (in units of $G_{0}$, the quantum
    of conductance) \textit{vs.} tunnel energy $E_{\mathrm{t}}$ in
    Co(0001) for P (a, bottom) and AP (b, top) alignment. The bias
    $E_{\mathrm{b}}$ ranges from $0.0$~eV (black) to $1.0$~eV (orange)
    and is indicated at the top of each dataset.}
  \label{fig:conductance2}
\end{figure}

When integrated over the 2BZ, one finds that $G(\mathrm{P}) >
G(\mathrm{AP})$ (cf. the black symbols in Fig.~\ref{fig:conductance2}
for zero bias).
With increasing bias, the conductances for tunneling at
$E_{\mathrm{F}}$ decrease. Since ballistic tunneling is a phase
coherent process, shifting of the electronic states of one electrodes
relative to those of the other by the bias, might reduce the phase
coherence. Or the spin-dependent DOS in the relevant region of the 2BZ
decreases with energy. We checked the spectral density carefully but
found no significant feature that would corroborate unequivocally the
latter explanation.

The conductances increase with tunnel energy $E_{\mathrm{t}}$ (cf.\ 
the data for $0.2$ to $1.0~\mathrm{eV}$ bias).  This fact might be
explained by the electronic structure of the leads or by a reduction
of the effective barrier width and height (cf.\ Figs.~\ref{fig:series}
and ~\ref{fig:barriershape}). Indeed, the radius of the ring-like
structure in the 2BZ which contributes most to the conductance (cf.\ 
Fig.~\ref{fig:transmission}) gets larger with tunnel energy, hence
increasing the contributing area. This is consistent with the focusing
effect mentioned earlier. The AP conductances in particular can be
represented reasonably well by parabolae.

A further interesting feature is the increase for biases of $0.6$,
$0.8$, and $1.0$~eV that occurs for P alignment at tunnel energies
$E_{\mathrm{t}}$ around $0.0$, $0.2$, and $0.4$~eV, respectively
(Fig.~\ref{fig:conductance2}a).  Inspection of the transmissions and
of the spectral density at $E_{\mathrm{F}} - 0.6~\mathrm{eV}$ produced
no significant feature that would explain this behavior (This is
corroborated by findings of LeClair \textit{et al.},
Ref.~\onlinecite{LeClair02}).  The feature occurs also for AP
alignment (Fig.~\ref{fig:conductance2}b) but not as pronounced as for
the P case.

\begin{figure}
  \centering
  \includegraphics[scale = 1.0]{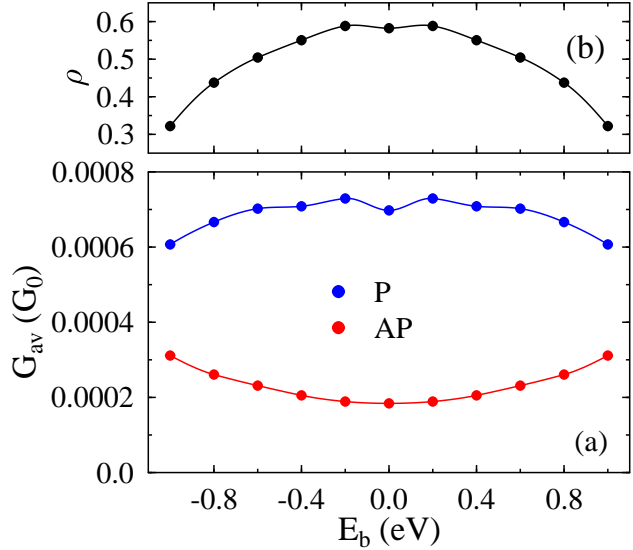}
  \caption{Magneto-resistance of vacuum tunneling in Co(0001).
    (a) Averaged conductance $G_{\mathrm{av}}$ (in units of $G_{0}$,
    the quantum of conductance) \textit{vs.} bias $E_{\mathrm{b}}$ for
    P (blue) and AP (red) alignment [cf.\ eq.~(\ref{eq:7})]. (b)
    Tunnel magneto-resistance $\rho$ \textit{vs.} $E_{\mathrm{b}}$
    [eq.~(\ref{eq:8})].}
  \label{fig:conductance}
\end{figure}

The increase with bias compensates the decrease for $E_{\mathrm{t}} =
E_{\mathrm{F}}$, as is shown for the averaged conductances
$G_{\mathrm{av}}$ in Fig.~\ref{fig:conductance}a.
Whereas $G_{\mathrm{av}}(\mathrm{P})$ decreases slightly (with a small
minimum at zero bias), $G_{\mathrm{av}}(\mathrm{AP})$ increases with
$|E_{\mathrm{b}}|$. Therefore, the resulting TMR $\rho$
[eq.~(\ref{eq:8})] drops with bias, too. However, the decrease which
is about $15~\%$ at $0.6~\mathrm{eV}$ is much less than that observed
for oxide barriers. In the latter case, the TMR drops by $50~\%$ to
$80~\%$ at $0.6~\mathrm{eV}$.~\cite{Yuasa02} Being due to the details
of the electronic structure in the Co leads, one could term the drop
in Fig.~\ref{fig:conductance}b as `DOS effect', rather than as
zero-bias anomaly.

Since inelastic processes (scattering at magnons, spin excitations)
are not included in our theory, one can conclude that the ZBA found in
tunnel junctions with oxide barriers can be attributed to defect
scattering in the oxide barrier. This finding is consistent with the
fact that the ZBA decreases with the improvement of the preparation
techniques for ferromagnet-oxide interfaces (see
Ref.~\onlinecite{Ding03a} and references therein).

In a previous investigation,~\cite{Ding03a} we used the superposition
approach (\ref{sec:superp-surf-barr}) for the tunnel barrier. There,
both the averaged conductances and the TMR were almost constant for
biases up to $0.5~\mathrm{eV}$. Comparing with the present results
that were obtained within the image-potential approach
(\ref{sec:image-charge-potent}), one has to keep in mind that details
of the calculations differ (e.\,g., the $\vec{k}_{\parallel}$ mesh).
However, these have only minor influence. The most striking difference
is the shape of the tunnel barrier which is varied in two aspects.
First, the JJJ barrier used in Ref.~\onlinecite{Ding03a} is rather
smooth with respect to the interpolating Lorentzian chosen here.
Generally speaking, the latter produces a larger reflection. Second,
the shape in the central part of the barrier differs. In particular
for large lead separations, the barrier height becomes important (cf.\ 
Figs.~\ref{fig:series} and \ref{fig:barriershape}). Further, the
linear bias potential which is missing in the superposition approach
is expected to have a non-negligible effect.  Therefore, details of
the barrier are expected to have significant influence on the tunnel
magneto-resistance.

\section{Concluding remarks}
\label{sec:concluding-remarks}
Tunnel calculations provide a rather indirect test of the proposed
barrier shapes. A more direct one would be to compare theoretical
energy positions and linewidths of so-called field-emission
resonances~\cite{Gadzuk93} with experimental ones. These electronic
states can be viewed as surface states that are trapped between the
bulk (in the presence of a bulk-band gap) and the tunnel barrier
between sample and an STM tip.  The field-emission resonances show up
as sharp maxima in the differential conductance and depend---like the
shape of the tunnel barrier---on both bias voltage and tip-sample
separation.

As a possible extension of the present work, one could think of a
treatment of tunnel junctions with `filled' spacers (instead of
vacuum), in particular with oxide barriers. Further, work is in
progress to describe the tunneling with bias voltage fully on an
\textit{ab-initio} level.

\acknowledgments
We would like to thank Hai Feng Ding, Arthur Ernst, Ingrid Mertig,
Silke Roether, Wulf Wulfhekel, and Peter Zahn for stimulating
discussions.

%
%
\bibliography{short,refs,refs2,refs3}

\begin{thebibliography}{42}
\bibitem[{Uni()}]{Units}
\bibinfo{note}{We use atomic Hartree units, $e = \hbar = m = 1$, $c \approx
  137.036$.}
\end{thebibliography}
\bibliographystyle{apsrev}

\end{document}